\documentclass[english,amsmath,amssymb,amsfonts,fleqn,10pt,floatfix,aps,superscriptaddress,twocolumn,nofootinbib,authoryear,tightenlines]{revtex4-1}

\usepackage[utf8]{inputenc}
\usepackage{lmodern}
\usepackage[english]{babel}
\usepackage{placeins}
\usepackage[usenames,dvipsnames]{color}
\usepackage{hyperref}
\usepackage{color}
\hypersetup{colorlinks,
linkcolor=Red,
citecolor=OrangeRed,
urlcolor=BlueViolet,
anchorcolor=OliveGreen}
\usepackage{booktabs}
\usepackage{relsize}
\usepackage{color}
\usepackage{units}
\usepackage{braket}
\usepackage{mathrsfs}
\usepackage{transparent}
\usepackage{verbatim}
\usepackage[us,12hr]{datetime}
\usepackage{filemod}
\usepackage[export]{adjustbox}
\usepackage{bm}
\usepackage{bbm}
\usepackage{dsfont}
\usepackage{pdfpages}
\usepackage{bibunits}
\defaultbibliography{DrivenSuperradiance}

\usepackage{microtype}
\usepackage{graphicx}
\newcommand{\mathsym}[1]{{}}
\newcommand{\unicode}[1]{{}}

\newcommand{\hide}[1]{}

\newcommand{\eq}[1]{Eq.\,(\ref{#1})}

\newcommand{\fig}[1]{Fig.\,\ref{#1}}

\newcommand{\half}[1]{\nicefrac{#1}{2}}
\newcommand{\brackets}[1]{\lbrace{#1\rbrace}}
\newcommand{\bp}{^\backprime}
\newcommand{\bpp}{^{\backprime\backprime}}

\newcommand*{\pdyad}[2]{ {\left(\ket{#1}\hspace{-0.5ex}\bra{#2}\right)}}

\newcommand*{\SpinBra}[2]{ \Bra{\hspace{-1ex}\begin{array}{l}#1\\#2\end{array}\hspace{-0.5ex}}}
\newcommand*{\SpinKet}[2]{ \Ket{\hspace{-0.5ex}\begin{array}{l}#1\\#2\end{array}\hspace{-1ex}}}
\newcommand{\nmap}[1]{{\operatorname{Map}_N{\mathlarger{\mathlarger{\left(\mathsmaller{\mathsmaller{ #1}}\right)}}}}}
\newcommand{\nmapalt}[0]{{\operatorname{Map}_N}}

\newcommand{\Or}{\mathcal{O}}

\newcommand {\ic}[0]{\ensuremath{\bm{i}}}
\newcommand {\id}[0]{\ensuremath{\mathds{1}}}
\newcommand {\jbar}[0]{\bar{J}}

\newcommand{\ox}[2]{{\operatorname{X}\limits_{#1}^{#2}}}

\definecolor{orange}{rgb}{0.8,0.4,0}

\begin{document}

\begin{bibunit}[apsrev4-1]
\setcounter{section}{0}
%\pagenumbering{arabic}

\title{Spin Squeezing by means of Driven Superradiance}
\author{Elie Wolfe}
\email{wolfe@phys.uconn.edu}
\affiliation{Department of Physics, University of Connecticut, Storrs, CT 06269}

\author{S.F. Yelin}
\affiliation{Department of Physics, University of Connecticut, Storrs, CT 06269}
\affiliation{ITAMP, Harvard-Smithsonian Center for Astrophysics, Cambridge MA 02138}

\date{\filemodprinttime{DrivenSuperradiance.tex}\hspace{1mm}on\hspace{1mm}\today}


\begin{abstract}
We discuss the possibility of generating spin squeezed states by means of driven superradiance, analytically affirming and broadening the finding in [\href[pdfnewwindow ]{http://link.aps.org/doi/10.1103/PhysRevLett.110.080502}{Phys. Rev. Lett. \textbf{110}, 080502 (2013)}]. In an earlier paper [\href[pdfnewwindow ]{http://link.aps.org/doi/10.1103/PhysRevLett.112.140402}{Phys. Rev. Lett. \textbf{112}, 140402 (2014)}] the authors determined that spontaneous purely-dissipative Dicke model superradiance failed to generate any entanglement over the course of the system's time evolution. In this article we show that by adding a driving field, however, the Dicke model system can be tuned to evolve toward an entangled steady state. We discuss how to optimize the driving frequency to maximize the entanglement. We show that the resulting entanglement is fairly strong, in that it leads to spin squeezing.% and the violation of Bell inequalities. We also discuss an extension of [Phys. Rev. Lett. \textbf{112}, 140402 (2014)] by considering alternative initial conditions as opposed to the canonical maximally excited state. Some alternative initial conditions lead to nonzero entanglement without the addition of a driving field.  
\end{abstract}
\maketitle

\section{Introduction}
In an earlier paper \cite{SuperradSeparable} the authors found that Dicke model superradiance \cite{superrad.original} did not generate entanglement. We show here, however, that entanglement can be generated in a multi-qubit system by means of {\em driven} superradiance, that is, when the system is additionally driven by some external field. Indeed, we qualitatively confirm the result of \citet{DrivenSuperradSpinPrecedent} that driven superradiance can be carefully tuned so as to generate spin squeezed states \cite{TothSpinSqueezing,CiracSpinSqueezing,SpinSqueezing2001,NoriSpinSqueezing}. Spin squeezed states are a class of entangled states which are particularly valuable for numerous specialized applications such as precision measurement \cite{SpinSqueezing2001,Gross2010Nature,NoriSpinSqueezing,PrecisionToth,Gross2012IOP}. In contrast to \citet{DrivenSuperradSpinPrecedent}, however, we here consider a measure of squeezing which is more generally more sensitive at detecting entanglement, and our findings therefore quantitatively differ; see Ref. \citep[Sec. 2]{NoriSpinSqueezing} and the Supplementary Online Materials for elaboration on this point. 

Our finding of spin squeezing in driven superradiance suggests that driven superradiance could potentially be a practical method for generating large entangled states \cite{SpinEntanglementMorrison,TothDickeArXiv,superrad2010,PhysRevA.83.013821}. Schema for generating entanglement in large systems are highly desirable, as they open the door for implementing quantum technologies such as information protocols which rely on a high bitrate of entangled qubits  \cite{DownConversion,ResonantFluorescence,PhysRevA.83.013821,PhysRevLett.109.173604}, such as for example, quantum key distribution \cite{CryptoPRA,CryptoPRL,RenatoQKDNature}. We here focus primarily on the spin squeezing measure for bipartite entanglement \cite{SpinSqueezing2001,NoriSpinSqueezing,TothSpinSqueezing,CiracSpinSqueezing} because of its especially versatile ramifications  \cite{SpinSqueezing2001,Gross2010Nature,NoriSpinSqueezing,PrecisionToth,Gross2012IOP}.

This driven superradiance scheme for generating spin squeezed states is particularly suitable for experimental implementation in that the entanglement generated is stable in time, by virtue of the system evolving toward an entangled steady state. We note that the system's entanglement never significantly exceeds its steady state entanglement value, as measured by the spin squeezing parameter, and thus there is no incentive to attempt careful pulse durations, which is experimentally very convenient. The system is also thoroughly robust, in the sense that the initial state of the system is irrelevant, as there is no bistability in the steady state solution. 

To be clear, the Dicke model of superradiance is the maximally simplified and idealized phenomenological model. We herein specifically study the Dicke model because it captures the essential fundamentals of superradiance while excluding confounding effects from consideration. The idealization employed in the Dicke model is that of perfect indistinguishability of the particles, such that we treat the system as existing entirely in only highest symmetry of the Hilbert space. Experimentally it corresponds to the small-volume limit and an absence of dipole-dipole induced dephasing. A realistic case, which we only briefly touch upon in this Letter, must account for dephasing and lower symmetry. A thorough treatment of the volume-dependent many-body effects not considered in the Dicke model can be found in Refs. \cite{superrad.yelinPRA,superrad.yelinBook}. We note that the driven variant of the Dicke model has been considered repeatedly, such as in Refs. \cite{DrivenSuperradCritPrecedent,PhysRevA.22.1179,PhysRevA.65.042107,DrivenSuperradSpinPrecedent}.

\section{Driven Superradiance Rate Equation}
Our system is modelled in Lindblad form by means of both a dissipative term as well as a driving potential. The dissipative term, expressed via Lindblad operators, corresponds to the spontaneous decay of the (open) system with decay rate $\Gamma$. We take the external driving frequency in our model to be $\omega$, and use the Rotating Wave Approximation \cite{scully1997quantum,RWADerivation1,RWADerivation2}. Thus, the Liouville master equation \citep[Eq. (2)]{Breuer2007Theory,ExactTwoLevel,PhysRevA.89.042120} which governs the time evolution of driven Dicke model superradiance is
\begin{align}\label{eq:lindblad}
&\hspace{-\mathindent}\frac{\partial \rho}{\partial t} =\Gamma\left(D^- \rho D^+ - \frac{D^+ D^- \rho + \rho D^+ D^-   }{2}\right) -\ic \left[V,\rho\right]
\end{align}
where 
\begin{align}\label{eq:potential}
    V &= \frac{\omega}{2}\left(D^+ + D^- \right)
\end{align}
and
\begin{align}\label{eq:Dplusminus}
    D^+ &= \sum\limits_{n=1}^N{\underbrace{\id ... \id}_{n-1}\otimes \pdyad{0}{1} \otimes \underbrace{\id ... \id}_{N-n}} %\\    D^- &= \sum\limits_{n=1}^N{\underbrace{\id ... \id}_{n-1}\otimes \pdyad{1}{0} \otimes \underbrace{\id ... \id}_{N-n}}
\end{align}
with the annihilation operator being the adjoint of the creation operator, $D^-=\left(D^+\right)^\dag$. Purely dissipative superradiance, such as is considered in Ref. \cite{SuperradSeparable}, is the special case of $\omega=0$.  We find that when the system is driven it tends towards a single steady state solution defined by $\dot \rho=0$.

%When the system is driven we find that it tends towards a single steady state solution defined by $\dot \rho=0$. It is convenient to change variables so as to work with a dimensionless time scale and a single dimensionless system-size-dependant driving parameter. We introduce
%\begin{align}\label{eq:OmegaDef}
%\tau=\Gamma t\,,\,\,\,\Omega = \omega/\Gamma/N\,,\text{ and }V^*=V/\Gamma\,,
%\end{align}
%transforming \eq{eq:lindblad} into
%\begin{align}\label{eq:lindblad2}
%&\hspace{-4ex}\frac{\partial \rho}{\partial \tau} =D^- \rho D^+ - \frac{D^+ D^- \rho + \rho D^+ D^-   }{2} -\ic \left[V^*,\rho\right]\,.
%\end{align}
To solve \eq{eq:lindblad} we need not consider a fully-general density matrix $\rho$. Firstly, the equation is symmetric with respect to permutation of the individual qubit Hilbert spaces, so we can take our density matrix to be symmetric, that is, expandable in symmetric basis states. Second, the raising and lowering nature of the driving potential allows us to infer which matrix elements must be real and which must be (entirely) imaginary, and therefore we can define a sufficiently-general $N$-particle density matrix
\begin{align}\begin{split}\label{eq:rhodef}
\hspace{-3ex}\rho_N&=\sum\limits_{m_a=-j}^{j}{\sum\limits_{m_b=-j}^{j}{\operatorname{X(j)}\limits_{m_a}^{m_b} {\ic}^{\left(m_a-m_b\right)}{\SpinKet{j}{m_a}\hspace{-0.8ex}\SpinBra{j}{m_b}}}}
\\&\text{ with real symmetric }\operatorname{X(j)}\limits_{m_a}^{m_b}=\operatorname{X(j)}\limits_{m_b}^{m_a} \in \mathbb{R} 
\end{split}\end{align}
using unnormalized Dicke states as our basis. Namely
\begin{align}\label{eq:unnorm}
%\SpinKet{j}{m}^{\vphantom{\ket{0}}}=\sum_{\begin{array}{c}\scriptstyle{\text{perms.}}\\ \scriptstyle{\brackets{\ket{0},\ket{1}}}\end{array}}{\ket{\underbrace{0...0}_{j-m},\underbrace{1...1}_{j+m}}}
\SpinKet{j}{m}=\operatorname{Symmetrize}\hspace{-0.5ex}{\Bigg[\,\;{\ket{\underbrace{0...0}_{j-m},\underbrace{1...1}_{j+m}}}\,\;\Bigg]}
\end{align}
and where 
\begin{align}\label{eq:jdef}
j=N/2
\end{align}is the total spin of our system of $N$ spin one-half particles. 
This basis is valuable because
\begin{align}
    \hspace{-\mathindent}&D^{\pm} \SpinKet{j}{m} = \begin{cases} (j \pm m+1)\SpinKet{j}{m \pm 1} & {\text{if }\;1\pm m \leq j} \\ 0 & \text{else.} \end{cases}
%    \\\hspace{-\mathindent}\text{and}\;&\SpinBra{j}{m} D^{\pm} = (j \mp m+1)\SpinBra{j}{m \mp 1} \mathbb{B}\left[1 \mp m \leq j\right]
\end{align}
%where $\bdelta{arg...}$ is a unitstep indicator function, returning one whenever its argument is true and zero otherwise\footnote{For example, $\bdelta{a\geq b}$ would be \texttt{Boole[a>=b]} in Mathematica{\it\tiny\texttrademark}. The unitstep indicator is somewhat analogous to the Heaviside Theta function, $\operatorname{\Theta}{\left(a-b\right)}$, but $\bdelta{a= b}\equiv 1$ when $a=b$.}, which we insert to indicate instances of annihilation of highest or lowest states. 
We can therefore express \eq{eq:lindblad} directly as a sum of the matrix elements $\ox{m_a}{m_b}$ over the summation indices $m_a$ and $m_b$. If we then re-index the dummy variables of summation so as to have a common index in the Dicke basis, as opposed to a common index in $\ox{m_a}{m_b}$, we obtain a set of coupled first-order differential equations defined by 
\begin{align}\begin{split}\label{eq:drivendiffeq}
&\hspace{-\mathindent}\frac{\partial\ox{m_a}{m_b}}{\partial t}=\Gamma \left(j-m_a\right) \left(j-m_b\right)\ox{m_a+1}{m_b+1}
\\&\hspace{-4ex}-\frac{ \Gamma}{2}\Big(\hspace{-0.5ex}\left(j-m_a+1\right) \left(j+m_a\right)%\bdelta{j+m_a\geq 1}
\\&\hspace{-0ex}+\left(j-m_b+1\right) \left(j+m_b\right) %\bdelta{j+m_b\geq 1}
\Big)\ox{m_a}{m_b}\\
&\hspace{-4ex}+ \frac{\omega}{2}\Big(\hspace{-0.5ex}\left(j-m_a\right) \ox{m_a+1}{m_b}+\left(j-m_b\right)\ox{m_a}{m_b+1}\\
&\hspace{-0ex}-\left(j+m_a\right)\ox{m_a-1}{m_b}-\left(j+m_b\right)\ox{m_a}{m_b-1}\Big)
\end{split}\end{align}
where we dropped various indicator functions by assuming that $-j\leq m_a,m_b \leq j$. %Note that we have also switched to dimensionless variables,
%\begin{align}\label{eq:OmegaDef}
%\tau=\Gamma t\text{ and }\Omega = \omega/\Gamma\,.
%\end{align}
\eq{eq:drivendiffeq} is our master rate equation; setting the left hand side to zero defines the steady state condition, along with
\begin{align}\label{eq:tracecondition}
\operatorname{tr}{\left[\rho_N\right]}=\sum\limits_{m=-j}^{j}{\binom{2j}{j+m}\operatorname{X(j)}\limits_{m}^{m}} =1
\end{align}
which is nontrivial only due to our choice of unnormalized basis states in defining the matrix elements, per \eq{eq:rhodef}. 

To obtain, practically, the steady-state matrix elements from \eq{eq:drivendiffeq} we need to iterate it over all possible $-j\leq m_a,m_b \leq j$, amounting to $(N+1)^2$ equations. Without loss of generality we can invoke the symmetry of the matrix elements to consider only $-j\leq m_a\leq m_b \leq j$, which reduces the set of equations by about a factor of two. Even leveraging the symmetry, however, the set of linear equations scales like  $\Or{\left(N^2\right)}$, and thus has quadratic computational complexity.

\section{The Spin Squeezing Parameter \({\xi}^2\)}
Spin Squeezing provides a valuable metric of entanglement \cite{SpinSqueezing2001,NoriSpinSqueezing,TothSpinSqueezing,CiracSpinSqueezing}, with extensive immediate application in precision metrology \cite{SpinSqueezing2001,Gross2010Nature,NoriSpinSqueezing,PrecisionToth,Gross2012IOP}. We use the explicit form of the spin squeezing parameter of \citet[Eq. (57)]{NoriSpinSqueezing} and \citet[Eq. (45)]{SpinSqueeze2013}, as follows:
\begin{align}\label{eq:chi}
&\hspace{-\mathindent}{\xi^2}=\frac{\Braket{{\jbar_1}^2+{\jbar_2}^2}-\sqrt{\Braket{{\jbar_1}^2-{\jbar_2}^2}^2+\Braket{{\jbar_1}{\jbar_2}+{\jbar_2}{\jbar_1}}^2}}{2/N}
\end{align}
where
\begin{align}\begin{split}\label{eq:j12}
    \jbar_1 &= \jbar_y\cos{\phi}-\jbar_x\sin{\phi}\\
    \jbar_2 &= \jbar_x\cos{\theta}\cos{\phi}+\jbar_y\cos{\theta}\sin{\phi}-\jbar_z\sin{\theta} 
\end{split}\end{align}    
and
\begin{align}\begin{split}\label{eq:phitheta}
    \theta &= \cos^{-1}{\left(\nicefrac{\Braket{\jbar_z}}{\sqrt{{\Braket{\jbar_x}}^2+{\Braket{\jbar_y}}+{\Braket{\jbar_z}}^2}}\right)} \\
    \phi &= \tan^{-1}{\left(\nicefrac{\Braket{\jbar_y}}{\Braket{\jbar_x}}\right)\,\text{, sensitive to quadrant,}}
\end{split}\end{align}
and where
\begin{align}
    \jbar_{x/y/z} &= \frac{1}{N}\sum\limits_{n=1}^N{\underbrace{\id ... \id}_{n-1}\otimes \sigma_{x/y/z} \otimes \underbrace{\id ... \id}_{N-n}} \,.
\end{align}

The calculation of \({\xi}^2\) can be immensely simplified by recognizing that the entire system's spin is encoded in the $\rho$'s one or two particle reduced states. For states with real and imaginary parts {\`a} la \eq{eq:rhodef} we show in the Supplementary Online Materials that
\begin{align}\label{eq:chiasmin}
&\hspace{-\mathindent}\xi^2=1+\left(N-1\right)\times\min \Bigg\lbrace\Braket{\sigma_x \otimes \sigma_x},
\\\nonumber &\hspace{-\mathindent}\frac{\Braket{\sigma_y}^2\Braket{\sigma_z \otimes \sigma_z}+\Braket{\sigma_z}^2\Braket{\sigma_y \otimes \sigma_y}-\Braket{\sigma_y}\Braket{\sigma_z}\Braket{\sigma_y \otimes \sigma_z}}{\Braket{\sigma_y}^2+\Braket{\sigma_z}^2}
%-\Braket{\sigma_y}\Braket{\sigma_z}\Braket{\sigma_y \otimes \sigma_z}}{\Braket{\sigma_y}^2+\Braket{\sigma_z}^2}
\Bigg\rbrace
\end{align}
For superradiating systems it is always the case that $\Braket{\sigma_x \otimes \sigma_x}$ is the more negative of the two terms. Using $\rho_N^{(d)}$ to indicate the reduced state of $d$ particles, and specifying $\rho$ in the expectation value purely for pedagogical clarity, we have
\begin{align}\label{eq:assumedmin}
\xi^2_{_N}&=1+\left(N-1\right)\Braket{\sigma_x \otimes \sigma_x \;\rho_N^{(2)}}\,,
\end{align}
for driven superradiance, where subscript $N$ indicates this special-case form. Note that a coordinate-independent generalization of \eq{eq:assumedmin} is well known to hold true for {\em all} symmetric states \citep[Eq. (7)]{SymmetricSpinSqueezing}.

$\xi^2_{_N}$ in \eq{eq:assumedmin} amounts to an upper-bound for the general spin-squeezing parameter for all $\rho$ of the form of \eq{eq:rhodef}. Since spin-squeezing is defined by ${\xi^2<1}$, with no loss of generality we therefore have certification of nonzero entanglement \cite{entang.review.toth,multireview,characterizingentanglement,SymmetricEquivalents,eckert2002quantum} via
\begin{align}\label{eq:guarantee}
\forall_N:\,\,\rho_N\in \brackets{\bm{\mbox{\large \(\varrho\)}}_\text{entangled}}\text{ if }\Braket{\sigma_x \otimes \sigma_x \;\rho_N^{(2)}}<0\,
\end{align}
which, since $\forall_N:\,\,\Braket{\sigma_x \;\rho_N^{(1)}}=0$, means that \eq{eq:guarantee} is just a special case of the the general entanglement criteria of \cite[Eq. (33)]{RamseySpinSqueezing,SymmetricSpinSqueezing,SymmetricEquivalentsPRL,OptimalSpinSqueezingParamater}, which recognizes that all separable symmetric states satisfy $\Braket{A \otimes A}-\Braket{A \otimes \id} \geq 0$ for all Hermitian operators A.

Now, for the two-particle state $\rho_2$ we find that the relevant expectation value is
\begin{align}\label{eq:sigma2}
\Braket{\sigma_x \otimes \sigma_x \;\rho_2} = 2\sum\limits_{s=0}^{1}{{\left(-1\right)}^{s}\operatorname{X(1)}\limits_{-s}^{s}}\,.
\end{align}
This is valuable because the two-particle reduced state for a general $\rho_N$ can be constructed by a  simple transformation acting on the matrix elements of $\rho_2$. Indeed, as shown in the Supplementary Online Materials, we have that $\rho_N^{(d)}=\nmapalt{\big(\rho_d\big)}$ where
\begin{align}\begin{split}\label{eq:mappingnew}
&\hspace{-3ex}\nmap{\operatorname{X(j\bp)}\limits_{m_a\bp}^{m_b\bp}}\to \sum\limits_{q=-\lambda}^{\lambda}{\binom{2 \lambda}{\lambda+q}\operatorname{X(j)}\limits_{m_a\bp+q}^{m_b\bp+q}}
\end{split}\end{align}
with $\lambda\equiv j-j\bp=\frac{N-d}{2}$, and therefore 
\begin{align}\label{eq:sigmamap}
&\Braket{\sigma_x \otimes \sigma_x \;\rho_N^{(2)}}=\nmap{\Braket{\sigma_x \otimes \sigma_x\;\rho_2^{\vphantom{(2)}}}}
\end{align}
such that \eq{eq:sigma2} allows us to explicitly express \eq{eq:assumedmin} as 
\begin{align}\label{eq:altchi}
&\hspace{-\mathindent}\xi^2_{_N}=1+2(N-1)\sum\limits_{s=0}^{1}{{\left(-1\right)}^{s}\hspace{-1ex}\sum\limits_{q=-\lambda}^{\lambda}{\binom{2\lambda}{\lambda+q}\operatorname{X(j)}\limits_{q-s}^{q+s}}}
\end{align}
using $\lambda=N/2-1$ to ``unpack" the two-particle expectation value of \eq{eq:sigma2}. This compact expression conveniently gives the spin squeezing parameter for a general $N$-particle driven superradiant state directly in terms of its matrix elements. Note that \eq{eq:altchi} is valid throughout the time evolution, and makes no assumptions about the system having obtained its steady state.

\section{Driving for Entanglement}
\begin{figure}[b]
\centering
%\begin{minipage}[t]{.48\textwidth}
\includegraphics[width=1\linewidth,center]{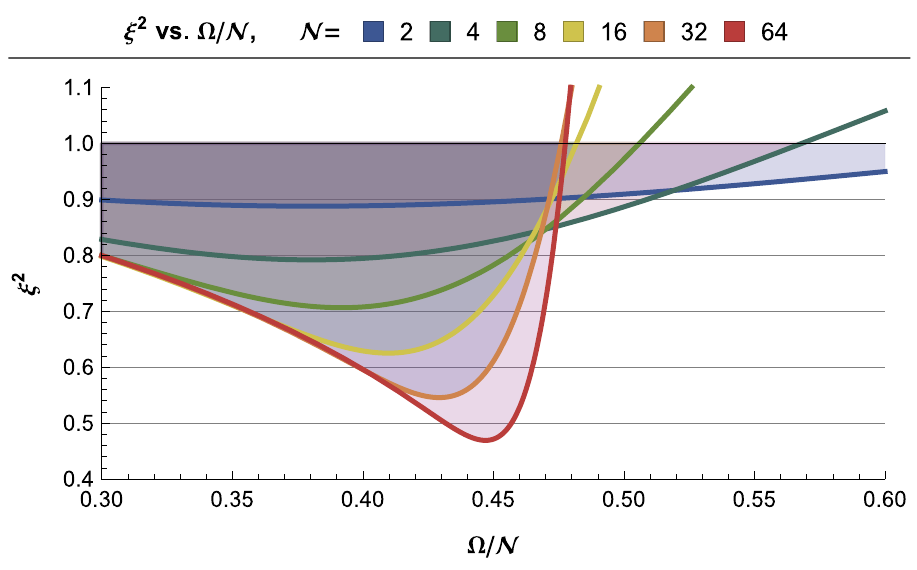}
\caption{(Color online) We graph  the spin-squeezing parameter $\xi^2$ for a steady state driven superradiant system as a function of $\Omega/N$ for various small $N$. Recall that $\Omega$ is the ratio of the driving frequency to relaxation frequency, $\Omega \equiv \omega/\Gamma$. The state is spin-squeezed whenever $\xi^2<1$; shown as shaded regions in this graph. The minima of the curves descends further with increasing $N$.}\label{fig:figXvO}
%\end{minipage}\hfill
\end{figure}

\begin{figure}[t]
%\centering
%\begin{minipage}[t]{.48\textwidth}
\includegraphics[width=0.99\linewidth,left]{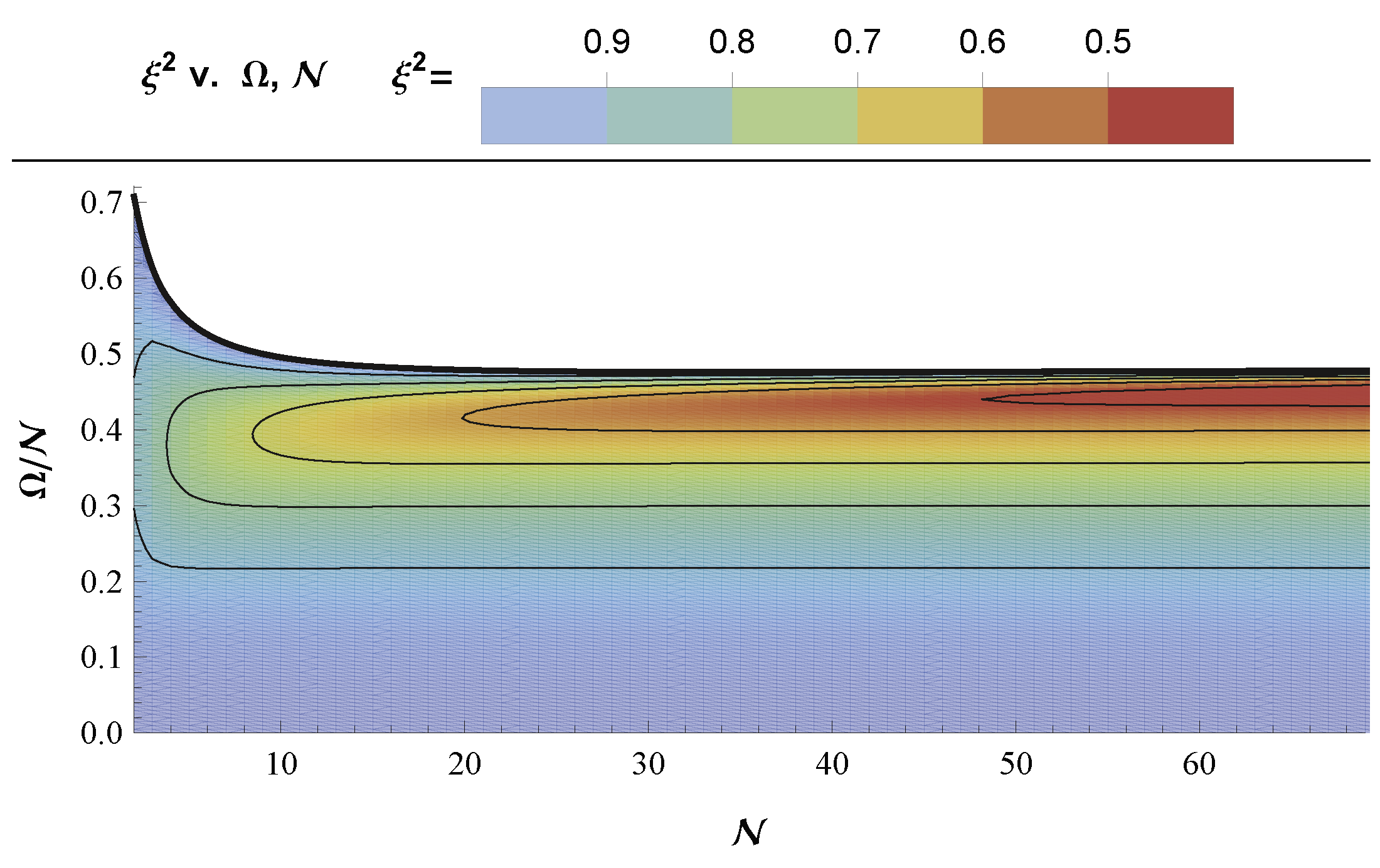}
\caption{(Color online) This contour plot shows the spin-squeezing parameter $\xi^2$ for a steady state driven superradiant system as a function of $\Omega = \omega/\Gamma$ over a dense set of $N$, among which are the discrete $N$ plotted in \fig{fig:figXvO}. %To emphasise the boundary of entanglement-inducing $\Omega$ 
We plot only the region where $\xi^2<1$. Red indicates strongest spin squeezing, ie. minimal $\xi^2<1$. Although hard to see, the $\xi^2=1$ boundary is {\em not} monotonically decreasing; rather, it's minimized at $\Omega/N=36$.}\label{fig:figOvN}
\end{figure}

Our question now is can we find some $\omega$ for a given $\Gamma$ such that we can drive the system into an entangled state characterized by $\xi^2<1$? Yes! We quantify the entanglement of the steady state in terms of $\Omega\equiv \omega/\Gamma\,$, defined as the ratio of the two experimental parameters. We find the steady state to be spin squeezed, ie. with measure $\xi^2<1$, for sufficiently small $\Omega$; see for example \fig{fig:figXvO}. To make a general statement, we note that for all $N$, when $ |\Omega| \lesssim 0.475 N$ the resulting steady state is always at least somewhat spin-squeezed state, see \fig{fig:figOvN}.

%\begin{minipage}[t]{.48\textwidth}

\begin{figure}[t]
\centering
\includegraphics[width=1\linewidth,center]{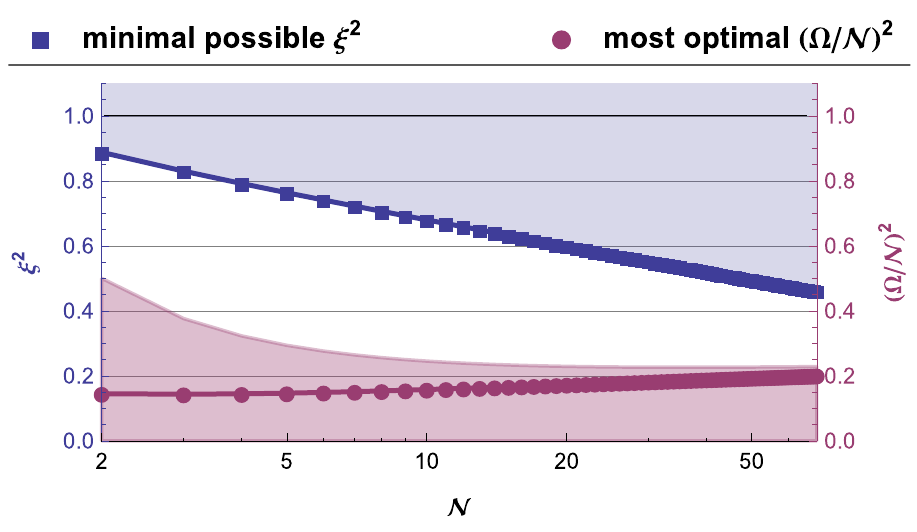}
\caption{(Color online) We plot the best-case scenario values for entanglement generation as a function of system size $N$. Note the dual meaning of the $Y$ axis: The upper curve indicates the minimal possible $\xi^2$, it is shaded upward to indicate that all larger values are also achievable. The lower curve indicates the optimal choice of $(\Omega/N)^2$ to achieve the corresponding minimal $\xi^2$. The lower shading indicates the complete parameter region where the steady state is spin squeezed. Note the logarithmic scaling of the $X$ axis.} % to highlight the asymptotically-logarithmic dependence of {\em both} the optimal $(\Omega/N)^2$ as well as the minimal $\xi^2$ on the system size $N$. }
\label{fig:figXvN}
\end{figure}

One would like to know how to tune $\Omega$ so as to maximize the entanglement in the resulting steady state. To this end, see \fig{fig:figXvN} where it appears that the optimal $\Omega$ scales like $(\Omega/N)^2 \sim a \ln{N}+b$ for large $N$. The precise dependence has not been clearly established, however; it is desideratum for future work. Although we found that  $\xi^2$ can be easily computed from less than $2N$ matrix elements of $\rho_N$ per \eq{eq:altchi}, obtaining those matrix elements requires solving  $\Or{\left(N^2\right)}$ linear equations, and therefore is not analytically amenable beyond small $N$. Some analytical results are tabulated in the Supplementary Online Materials.

It is also desirable to quantify the maximal extent of the spin squeezing that can achieved in the our model of driven Dicke superradiance. Per \fig{fig:figXvN}, we observe a rapid strengthening of the squeezing extent as we consider larger systems. Indeed, the value of the best-possible $\xi^2$ almost appears to drop off logarithmically as a function of $N$, descending below 0.5 at the right edge of \fig{fig:figXvN} with no sign yet of tapering off. This suggest that by increasing the size of the system, $\xi^2$ can perhaps be made arbitrarily small in the steady state of our model. With the usual caveats that genuine superradiance suffers from volume-dependant effect not accounted for in the Dicke model \cite{superrad.yelinPRA,superrad.yelinBook}, this result nevertheless further suggest that driven superradiance may be a viable scheme for generating large tightly squeezed states.

%\begin{figure*}[t]
%\centering
%\begin{minipage}[t]{.48\textwidth}
%\includegraphics[width=0.9\linewidth,left]{SmallWolfePorrasComparison.pdf}
%\caption{This graph shows the spin-squeezing parameter $\xi^2$ for a steady state driven superradiant system as a function of the $\Omega$ for various small $N$. Recall that $\Omega$ is the ratio of the driving frequency to relaxation frequency, $\Omega = \omega/\Gamma$ per \eq{eq:OmegaDef}. The state is spin-squeezed whenever $\xi^2<1$ or $1/\xi^2>1$. Observe that optimal spin squeezing is achieved at $|\Omega|\sim 0.4$.}\label{fig:fig27}
%\end{minipage}\hfill
%\begin{minipage}[t]{.48\textwidth}
%\includegraphics[width=0.9\linewidth,left]{LargeWolfePorrasComparison.pdf}
%\caption{This graph extends \fig{fig:fig27} to larger $N$ up to $N=30$. The steady-state solution is computed from a system of $\Or{\left(N^2\right)}$ linear equations; this quadratic scaling makes finding the steady-state $\rho$ prohibitive for larger $N$. Note that optimal spin squeezing appears to tend towards $|\Omega|\sim 1/2$.}\label{fig:fig530}
%\end{minipage}
%\end{figure*}

\section{Negativity}
A well known necessary condition for separability is that $\rho$ should be positive semidefinite under all possible partial transpositions \cite{PPTAsher,PPTHorodecki}. If, under the transposition of some Hilbert subspace, $\rho^{PT}$ is found to have one or more negative eigenvalues, then $\rho$ is known to be entangled. The entanglement monotone {\em Negativity}.\cite{OneNegativeEigenvalue,NegativityIntro,multireview,PlenioEntanglementMeasures} is equal to the combined magnitude of the negative eigenvalues, ie.
\begin{align}\label{eq:negativity}
\operatorname{\mathscr{N}}{\hspace{-0.25ex}\Big(\rho\Big)} &\equiv \frac{\big(\sum_{i}{\left|\lambda_i\right|}\big)-1}{2}\,.
\end{align}
The Negativity is a common benchmark of a state's distillability and resource value for nonlocality \cite{BellViolation,GivenQB.Cabello}. 

For a $2\times 2$ system, such as $\rho_N^{(2)}$, it is known that the partial transpose is always full rank and has at most one negative eigenvalue \cite{OneNegativeEigenvalue}, in which case the Negativity is the magnitude of that single negative eigenvalue. By direct computation we find that $\nicefrac{\Braket{\sigma_x \otimes \sigma_x \;\rho_2^{\vphantom{(2)}}}}{2}$ is one of the eigenvalues of $\rho_2^{PT}$, and thus via the mapping of \eq{eq:sigmamap} we also have that $\nicefrac{\Braket{\sigma_x \otimes \sigma_x \;\rho_N^{(2)}}}{2}$ must be an eigenvalue of general ${\rho_N^{(2),PT}}$. What we see is that the spin-squeezing parameter is effectively a linear function of the reduced state Negativity, such that \eq{eq:assumedmin} has the corollary 
\begin{align}\label{eq:negrelation}
\hspace{-4ex}\text{If }\xi^2_{_N}<0\,,\text{ then }\;\xi^2_{_N} & = 1-2\left(N-1\right)\operatorname{{\mathscr{N}}}{\hspace{-0.25ex}\Big(\rho_N^{(2)}\Big)}\,.
\end{align}
See Refs. \cite{OneNegativeEigenvalue,PhysRevA.70.022322} for a translation between the Negativity and Concurrence entanglement monotones, as the Concurrence has in some sense become a conventional standard metric for multiparticle entanglement \cite{PhysRevLett.95.260502}, such as in Refs. \cite{SpinEntanglementMorrison,PhysRevA.82.012327}. Spin squeezing is directly related to the two-particle Concurrence in Ref. \citep[Eq. (5)]{RamseySpinSqueezing} and to the CCNR criteria in Ref. \citep[Obs. 2]{SymmetricEquivalents}.

\section{Outlook}
We have shown analytically and numerically that driven Dicke model superradiance leads to temporally-stable entangled states with nontrivial spin squeezing parameter, as was first noted by  \citet{DrivenSuperradSpinPrecedent}. We consider the essential novel contributions in this Letter to be the derivation of the rate equation for driven superradiance directly in terms of well-defined matrix elements [\eq{eq:drivendiffeq}], and the expression of the spin squeezing parameter also directly in terms of such elements [\eq{eq:altchi}]. Together this allows for computationally optimal computation of $\xi^2$, without requiring matrix multiplication and with minimal memory overhead. Our final results are quantitatively different from those of \citet{DrivenSuperradSpinPrecedent} only since we elected to use an alternative measure for spin squeezing, one which is more sensitive at detecting entangled states, as discussed in the Supplementary Online Materials. We also explored somewhat how to optimize the steady state entanglement [\fig{fig:figXvN}]. 

We emphasize that not only is the entanglement in driven Dicke model superradiance invariant in time, and insensitive to initial conditions, we furthermore observe that the extent of the squeezing appears to be unlimited as the system size scales up. This encouraging result all the more suggests that genuine driven superradiance may be a potentially viable scheme for practical large-scale entanglement generation. 

Because the Dicke model represents the extreme ideal limit, it is therefore desirable to further consider a model which more closely represents experimentally achievable phenomena, so as to better assess the realistic candidacy of driven superradiance for generating entanglement. Refs. \cite{superrad.yelinPRA,superrad.yelinBook} introduce ways to calculate superradiant dynamics for a much more realistic case. Preliminary numeric calculations of driven superradiance per that context continue to indicate the existence of range of $\omega$ such that the corresponding steady state is spin squeezed. The persistence of this defining qualitative feature even in a realistic model suggest that the spin squeezing properties discussed in this Letter hold in a similar manner in the presence of interactions and finite dephasing. In a forthcoming paper we will discuss this realistic case along with explicit examples of superradiant squeezing improved measurements of clock and spin systems.

We are grateful to Dr. Diego Porras of the University of Sussex for valuable discussions and to Dr. Géza Tóth of the Hungarian Academy of Sciences for bringing to our attention the spin squeezing parameter which is most optimal for detecting entanglement. We wish to thank the NSF and the AFOSR for funding.

%\; \color{orange} This is a somewhat too weak and rather academic way to say this; may I suggest something more along those lines: those references introduce ways to calculate superradiant dynamics for a much more realistic case. Preliminary calculations of the squeezing parameter in that context suggest that the spin squeezing properties discussed in this Letter hold in a similar manner in the presence of interactions and finite dephasing. In a forthcoming paper, we will discuss this realistic case along with explicit examples of superradiant squeezing improved measurements of clock and spin systems.\color{black}  we intend to present those results in a forthcoming paper.

%\FloatBarrier
%\vspace*{-4ex}
\putbib
\end{bibunit}
\bibliographystyle{apsrev4-1}
%\vspace*{-4ex}\bibliography{DrivenSuperradiance}

\begin{bibunit}[alphaurl]

\onecolumngrid
\clearpage
\appendix
%\section{APPENDICES}
\renewcommand{\theequation}{S\arabic{equation}}

\section{Derivation of the reduced state form}
Another benefit of using the unnormalized Dicke states of \eq{eq:unnorm} in defining our matrix elements in \eq{eq:rhodef} is that they enable us to partition the Hilbert space without requiring Clebsch-Gordon coefficients. That is to say,
\begin{align}
\SpinKet{J}{M}=\sum\limits_{m=-j}^{j}{\SpinKet{J-j}{M-m}\otimes\SpinKet{j}{m}}
\end{align}
or, rather conveniently, we can split the $N=d+\kappa$ qubits in our very definition of $\rho$, \eq{eq:rhodef}. Taking $j=N/2$ as per \eq{eq:jdef} and defining from here on out $j\bp  = d/2$ and $j\bpp = \kappa/2$ 
%\begin{align}\label{eq:jdefs}
%j=N/2\text{  and  }j\bp  = d/2\text{  and  }j\bpp = \kappa/2
%\end{align}
we may re-express \eq{eq:rhodef} as
\begin{align}\begin{split}\label{eq:rhosplit}
\rho_N=&\sum\limits_{m_a\bp=-j\bp}^{j\bp}\sum\limits_{m_a\bpp=-j\bpp}^{j\bpp}\sum\limits_{m_b\bp=-j\bp}^{j\bp}\sum\limits_{m_b\bpp=-j\bpp}^{j\bpp} \left( \operatorname{X(j)}\limits_{m_a\bp+m_a\bpp}^{m_b\bp+m_b\bpp} {\ic}^{\left(m_a\bp+m_a\bpp-m_b\bp-m_b\bpp\right)}\times  \SpinKet{j\bp}{m_a\bp}\otimes\SpinKet{j\bpp}{m_a\bpp}\SpinBra{j\bp}{m_b\bp}\otimes\SpinBra{j\bpp}{m_b\bpp} \right)
\end{split}\end{align}
This allows us to compute reduced states $\rho^{(d)}$ where all but $d$ particles out of the full $N$ have been traced out. Tracing out $\kappa$ particles means 
\begin{align}
&\rho_N^{(d)}=\underbrace{\sum\limits_{i_1=0}^{1}...\sum\limits_{i_\kappa=0}^{1}}_{\kappa=N-d}{\left(\id\otimes \bra{i_1}...\bra{i_\kappa}\right)} \rho_N {\left(\id\otimes \ket{i_1}...\ket{i_\kappa}\right)}
\end{align}
where because in the unnormalized Dicke basis of \eq{eq:unnorm} we rather simply have
\begin{align}
\bra{i_0}...\bra{i_\kappa} \cdot \SpinKet{j\bpp}{m\bpp} =\begin{cases} 1 & \text{if }\;j\bpp+m\bpp=\sum\limits_{s=1}^{\kappa}{i_s} \\ 0 & \text{else} \end{cases}
%\bdelta{j\bpp+m\bpp=\sum\limits_{s=1}^{\kappa}{i_s}}
\end{align}
and therefore we can recognize that
\begin{align}\begin{split}
 \sum\limits_{i_1=0}^{1}...&\sum\limits_{i_\kappa=0}^{1}{\left(\id^{\otimes d}\otimes \bra{i_1}...\bra{i_\kappa}\right)} f[m_a\bp+m_a\bpp]\SpinKet{j\bp}{m_a\bp}\otimes\SpinKet{j\bpp}{m_a\bpp}
\\&=\SpinKet{j\bp}{m_a\bp}{\left( \sum\limits_{i_1=0}^{1}\bra{i_1}...\sum\limits_{i_\kappa=0}^{1}\bra{i_\kappa}\right)} {\SpinKet{j\bpp}{m_a\bpp}f[m_a\bp+m_a\bpp]} 
\\&=\SpinKet{j\bp}{m_a\bp}\sum\limits_{i=0}^{\kappa}{\binom{\kappa}{i}\delta{\left(i=j\bpp+m_a\bpp\right)}f[m_a\bp+m_a\bpp]}
\\&=\SpinKet{j\bp}{m_a\bp}\sum\limits_{i=0}^{\kappa}{\binom{\kappa}{i}\delta{\left(m_a\bpp=i-\half{\kappa}\right)}f[m_a\bp+m_a\bpp]}
\end{split}\end{align}
\begin{comment}
making the change of variable  $ m_a=M-m$\,,
\begin{align}\begin{split}
&\hspace{-\mathindent}\SpinKet{\frac{N-\kappa}{2}}{m_a}{\left( \sum\limits_{i_1=0}^{1}\bra{i_1}...\sum\limits_{i_\kappa=0}^{1}\bra{i_\kappa}\right)}\hspace{-1ex} \sum\limits_{m=\frac{-\kappa}{2}}^{{\kappa}/2}{\hspace{-1ex}f[m_a+m]\SpinKet{\half{\kappa}}{m}}
\\&\hspace{-\mathindent}=\SpinKet{d/2}{m_a}\sum\limits_{m=\frac{-\kappa}{2}}^{{\kappa/2}}\sum\limits_{i=0}^{\kappa}\binom{\kappa}{i}\delta{\left(i=\half{\kappa}+m\right)}f[m_a+m]
\\&\hspace{-\mathindent}=\SpinKet{d/2}{m_a}\sum\limits_{i=0}^{\kappa}\binom{\kappa}{i}f[m_a+i-\half{\kappa}]
\end{split}\end{align}
\end{comment}
and therefore we can reduce states of the form of \eq{eq:rhodef}, namely
\begin{align}\label{eq:premapping}\begin{split}
%\hspace{-\mathindent}&\begin{array}{rlr}
{\rho_N^{(d)}=} & {\sum\limits_{m_a\bp=-j\bp}^{j\bp}{\sum\limits_{m_b\bp=-j\bp}^{j\bp}{\sum\limits_{i=0}^{\kappa}}}\left({{{\binom{\kappa}{i}{\operatorname{X(j)}\limits_{m_a\bp+i-\half{\kappa}}^{m_b\bp+i-\half{\kappa}} {\ic}^{\left(m_a\bp-m_b\bp\right)}{\SpinKet{j\bp}{m_a\bp}\hspace{-0.8ex}\SpinBra{j\bp}{m_b\bp}}}}}}\right)} \quad\text{ or, equivalently,}
%\end{split}\end{align}
%or equivalently 
%\begin{align}\label{eq:premapping1}\begin{split}
%\hspace{-\mathindent}&\begin{array}{rlr}
\\{\rho_N^{(d)}=} & {\sum\limits_{m_a\bp=-j\bp}^{j\bp}{\sum\limits_{m_b\bp=-j\bp}^{j\bp}{\sum\limits_{q=-\lambda}^{\lambda}}}\left({{{\binom{2 \lambda}{\lambda+q}{\operatorname{X(j)}\limits_{m_a\bp+q}^{m_b\bp+q} {\ic}^{\left(m_a\bp-m_b\bp\right)}{\SpinKet{j\bp}{m_a\bp}\hspace{-0.8ex}\SpinBra{j\bp}{m_b\bp}}}}}}\right)} 
\end{split}\end{align}
with $\lambda=\kappa/2=j\bpp=j-j\bp=\frac{N-d}{2}$. Note how the reduced $\rho_N^{(d)}$ has a strikingly similar form to \eq{eq:rhodef}, to the extent that we can simply summarize
\begin{align}\begin{split}\label{eq:mapping}
\rho_N^{(d)}=\nmapalt{\big(\rho_d\big)}\,,\;\text{ where 
}\;\nmap{\operatorname{X(j\bp)}\limits_{m_a\bp}^{m_b\bp}}\to \sum\limits_{q=-\lambda}^{\lambda}{\binom{2 \lambda}{\lambda+q}\operatorname{X(j)}\limits_{m_a\bp+q}^{m_b\bp+q}}
\end{split}\end{align}
%\begin{align}\begin{split}\label{eq:mapping}
%\rho_N^{(d)}=  \,\rho&_{j\bp=d/2} \,\text{ transformed such that } \,\operatorname{X(j\bp)}\limits_{m_a\bp}^{m_b\bp} \rightarrow \sum\limits_{q=-\lambda}^{\lambda}{\binom{2 \lambda}{\lambda+q}\operatorname{X(j)}\limits_{m_a\bp+q}^{m_b\bp+q}} 
%\end{split}\end{align}
and thus reduced states have precisely the same form as $N$ particle states, with the mapping between parameters given by \eq{eq:mapping} pursuant to \eq{eq:jdef}. This is an essential element we draw upon in constructing a simplified expression for the spin squeezing parameter.
%\centering
%\clearpage\includegraphics[page=1,width=1\textwidth]{CleanMasterEqnProof.pdf}
%\clearpage\includegraphics[page=2,width=1\textwidth]{CleanMasterEqnProof.pdf}
%\clearpage\includegraphics[page=3,width=1\textwidth]{CleanMasterEqnProof.pdf}

%\clearpage\includepdf[pages=1]{CleanMasterEqnProof2.pdf}
%\clearpage\includepdf[pages=2]{CleanMasterEqnProof2.pdf}

\section{Derivation of the form of the spin squeezing parameter}

When we specialize to symmetric states with real and imaginary structure consistent as per \eq{eq:rhodef} we find that great simplification of the spin squeezing parameter is possible. Note, for example, that 
\begin{align}\begin{split}\label{eg:expecvalueintro}
    \Braket{\jbar_{x/y/z}} &= \operatorname{tr}{\left[\sigma_{x/y/z} \otimes \underbrace{\id ... \id}_{N-1}\cdot \rho_N\right]} = \operatorname{tr}{\left[\sigma_{x/y/z} \cdot \rho_N^{(1)}\right]} \equiv \Braket{\sigma_{x/y/z}\;\rho_N^{(1)}}\,. \end{split}
\end{align}
Furthermore, we can see that 
\begin{align}\label{eq:traceless1}
&\forall_N:\,\,\Braket{\sigma_x\;\rho_N^{(1)}}=0
\end{align}
as a consequence of 
\begin{align}\label{eq:nosigmax}
&\Braket{\sigma_x\;\rho_1}=0
\end{align}
and $\nmapalt{\big(0\big)}=0$ via \eq{eq:mapping}. This now implies that per \eq{eq:phitheta} we can identify
\begin{align}\begin{split}\label{eq:ourcoordinates}
    \sin{\phi}&=\operatorname{sgn}{\left(\Braket{\sigma_y}\right)},\quad \cos{\phi}=0,
    \quad \cos{\theta}\sin{\phi} = \frac{\operatorname{sgn}{\left(\Braket{\sigma_y}\right)}\Braket{\sigma_z}}{\sqrt{\Braket{\sigma_y}^2+\Braket{\sigma_z}^2}},
    \quad \text{and   }\sin{\theta} = \frac{\operatorname{sgn}{\left(\Braket{\sigma_y}\right)}\Braket{\sigma_y}}{\sqrt{\Braket{\sigma_y}^2+\Braket{\sigma_z}^2}}\end{split}
\end{align}
and therefore
\begin{align}
    &\jbar_1 = \frac{-\operatorname{sgn}{\left(\Braket{\sigma_y}\right)}\left(\sum\limits_{n=1}^N{\underbrace{\id ... \id}_{n-1}\otimes \sigma_{x} \otimes \underbrace{\id ... \id}_{N-n}}\right)}{N}\quad\text{and} \\
    &\jbar_2 = \frac{\operatorname{sgn}{\left(\Braket{\sigma_y}\right)}\Braket{\sigma_z}\left( \sum\limits_{n=1}^N{\underbrace{\id ... \id}_{n-1}\otimes \sigma_y \otimes \underbrace{\id ... \id}_{N-n}}\right)-\operatorname{sgn}{\left(\Braket{\sigma_y}\right)}\Braket{\sigma_y}\left( \sum\limits_{n=1}^N{\underbrace{\id ... \id}_{n-1}\otimes \sigma_z \otimes \underbrace{\id ... \id}_{N-n}}\right) }{N\sqrt{\Braket{\sigma_y}^2+\Braket{\sigma_z}^2}}
\end{align}
which can now be leveraged even further. By permutation symmetry one can see that
%\begin{align}\Braket{{\jbar_1}^2} = &\frac{\operatorname{tr}{\left[ {\sigma_x}^2 \cdot \rho^{(1)} \right]} +(N-1)\operatorname{tr}{\left[ {\sigma_x}\otimes{\sigma_x} \cdot \rho^{(2)} \right]}}{N}
%\end{align}
%which because ${\sigma_{x/y/z}}^2=\id$ simplifies to 
\begin{align}\label{eq:j1som}\Braket{{\jbar_1}^2} = &\frac{1}{N}+\left(\frac{N-1}{N}\right)\Braket{\sigma_x \otimes \sigma_x}\,.
\end{align}
Similarly we find that 
\begin{align}\label{eq:j2som}\Braket{{\jbar_2}^2} = &\frac{1}{N}+\left(\frac{N-1}{N}\right)\frac{\Braket{\sigma_y}^2\Braket{\sigma_z \otimes \sigma_z}+\Braket{\sigma_z}^2\Braket{\sigma_y \otimes \sigma_y}-\Braket{\sigma_y}\Braket{\sigma_z}\Braket{\sigma_y \otimes \sigma_z}}{\Braket{\sigma_y}^2+\Braket{\sigma_z}^2}
%\Braket{\sigma_y}^2+\Braket{\sigma_z}^2-(N-1)\Braket{\sigma_y}\Braket{\sigma_z}\Braket{\sigma_y \otimes \sigma_z}}{N{\left(\Braket{\sigma_y}^2+\Braket{\sigma_z}^2\right)}}\,.
\end{align}
%\begin{align}
%&\Braket{{\jbar_2}^2} = \frac{\operatorname{tr}{\left[ {\left(\Braket{\sigma_z}\sigma_y-\Braket{\sigma_y}\sigma_z \right)}^2 \cdot \rho^{(1)} \right]} }{N{\left(\Braket{\sigma_y}^2+\Braket{\sigma_z}^2\right)}}
%\\\nonumber&\hspace{-\mathindent}+\frac{\operatorname{tr}{\left[ \left(\Braket{\sigma_z}\sigma_y-\Braket{\sigma_y}\sigma_z \right)\otimes \left(\Braket{\sigma_z}\sigma_y-\Braket{\sigma_y}\sigma_z \right) \cdot \rho^{(2)} \right]} }{N{\left(\Braket{\sigma_y}^2+\Braket{\sigma_z}^2\right)}{(N-1)^{-1}}}\,.
%\end{align}
%Since $\operatorname{tr}{\left(\sigma_y \sigma_z \rho^{%(1)} \right)}=\operatorname{tr}{\left(\sigma_z \sigma_y \rho^{(1)} \right)}=0$ the first term is just $1/N$ and thus
%\begin{align}
%&\Braket{{\jbar_2}^2} = \frac{1}{N}+\\\nonumber&\hspace{-\mathindent}\frac{\Braket{\sigma_z}^2\Braket{\sigma_y \otimes \sigma_y}-2\Braket{\sigma_y}\Braket{\sigma_z}\Braket{\sigma_y \otimes \sigma_y}+\Braket{\sigma_y}^2\Braket{\sigma_z \otimes \sigma_z} }{N\left(\Braket{\sigma_y}^2+\Braket{\sigma_z}^2\right){(N-1)^{-1}}}
%\end{align}
Lastly, we have that
\begin{align}\label{eq:zeroj}
&\forall_N:\,\,\Braket{\jbar_1 \jbar_2+\jbar_2 \jbar_1\;\rho_N^{(2)}} = 0
\end{align}
due to the readily-verifiable properties
\begin{align}\label{eq:traceless2}
\Braket{\sigma_{y}\otimes\sigma_{x} \: \rho_2}=0\,,\quad\Braket{\sigma_{z}\otimes\sigma_{x} \: \rho_2}=0\,
\end{align}
again leveraging $\nmapalt{\big(0\big)}=0$.

Note that \eq{eq:zeroj} immediately simplifies 
%\frac{\operatorname{tr}{\left[ \left(\Braket{\sigma_z}\sigma_y-\Braket{\sigma_y}\sigma_z \right)\sigma_x \cdot \rho^{(1)} \right]} }{N{\left(\Braket{\sigma_y}^2+\Braket{\sigma_z}^2\right)}}
%\\\nonumber&+\frac{(N-1)\operatorname{tr}{\left[ \left(\Braket{\sigma_z}\sigma_y-\Braket{\sigma_y}\sigma_z \right)\otimes\sigma_x \cdot \rho^{(2)} \right]} }{N{\left(\Braket{\sigma_y}^2+\Braket{\sigma_z}^2\right)}}
%\end{split}\end{align}
%but happily for states with real and imaginary parts {\`a} la \eq{eq:rhodef} we find that 
%\begin{align}\label{eq:traceless2}
%\operatorname{tr}{\left[\sigma_{y}\otimes\sigma_{x} \cdot \rho^{(2)}\right]}=\operatorname{tr}{\left[\sigma_{z}\otimes\sigma_{x} \cdot \rho^{(2)}\right]}=0
%\end{align}
%along the same lines as the special result of \eq{eq:traceless1}.
%Moreover note that $\sigma_{y/z}\otimes\sigma_{x} = -\sigma_{x} \otimes \sigma_{y/z}$ due to anticommutation, and as such 
%\begin{align}
%\Braket{\jbar_1 \jbar_2+\jbar_2 \jbar_1} = 0
%\end{align}
\eq{eq:chi} into
\begin{align}
\xi^2_{_N}&=\frac{N}{2}\left(\Braket{{\jbar_1}^2+{\jbar_2}^2}-\sqrt{\Braket{{\jbar_1}^2-{\jbar_2}^2}^2}\right)\,\text{ or, }\,
\xi^2_{_N}=N \min{\Big\lbrace\Braket{{\jbar_1}^2},\;\Braket{{\jbar_2}^2}\Big\rbrace}\,,
\end{align}
thus leading to \eq{eq:chiasmin} in the main text.

\clearpage
\section{Contrasting definitions of the Spin Squeezing Parameter}

In the main text we used the spin squeezing parameter $\xi^2$ as defined in \eq{eq:chi}, which we noted can be found in Refs. \citep[Eq. (57)]{NoriSpinSqueezing} and \citep[Eq. (45)]{SpinSqueeze2013}. The measure is denoted with a subscript $S$ in Ref. \cite{NoriSpinSqueezing}, where it is credited to Ref. \citep{KitagawaUeda} and is equivalently defined as
\begin{align}
&{\xi^2_S}=\frac{4}{N}\min\left(\Delta J_{\vec{n}_{\perp}}^{2}\right)
\end{align}
where $J_{\vec{n}_{\perp}}$ refers to the spin measured along some direction orthogonal to the mean spin, and the minimization is performed over all the vectors that lie in the plane orthogonal to the mean spin. This minimization is accounted for in the explicit formulation we used in \eq{eq:chi}. 

Using $\phi$ and $\theta$ as per \eq{eq:phitheta}, or more specifically for us, as per \eq{eq:ourcoordinates}, one can define the optimal direction of ${\vec{n}_{\perp}}$ \citep[Eq. (50)]{NoriSpinSqueezing} as 
\begin{align}
\vec{n}_{\perp}= \{\cos (\theta ) \sin (\varphi ) \cos (\phi )-\cos (\varphi ) \sin (\phi ),\cos (\theta ) \sin (\varphi ) \sin (\phi )+\cos (\varphi ) \cos (\phi ),-\sin (\theta ) \sin (\varphi )\}
\end{align}
with
\begin{align}
\varphi=\left\{ \begin{array}{ll}
\frac{1}{2}\arccos\Big(\frac{-A}{\sqrt{A^{2}+B^{2}}}\Big) & \text{if}~~B\leq0,\\
\pi-\frac{1}{2}\arccos\Big(\frac{-A}{\sqrt{A^{2}+B^{2}}}\Big) & \text{if}~~B>0,\end{array}\right.\
\end{align}
with
\begin{align}
A\equiv \Braket{{\jbar_1}^2-{\jbar_2}^{2}},\text{ \ }\; B\equiv \frac{\Braket{\jbar_1 \jbar_2+\jbar_2 \jbar_1}}{2}.
\end{align}
For the states we are considering, $B=0$ per \eq{eq:zeroj}. As effectively noted following \eq{eq:chiasmin}, we observe that $\Braket{{\jbar_1}^2}<\Braket{{\jbar_2}^{2}}$; see Eqs. (\ref{eq:j1som},\ref{eq:j2som}). The end result is that $\vec{n}_{\perp}=\hat{x}$.

In contrast, the spin squeezing parameter used in Ref. \cite{DrivenSuperradSpinPrecedent} is distinct from our formulation; their parameter is denoted in Ref. \citep[Eq. (80)]{NoriSpinSqueezing} by a subscript $R^{\prime}$, namely
\begin{align}\label{eq:xiconventional}
\xi_{R^{\prime}}^{2}=\frac{N\left(\Delta
J_{\vec{n}_{1}}\right)^{2}}{\left\langle
J_{\vec{n}_{2}}\right\rangle ^{2}+\left\langle
J_{\vec{n}_{3}}\right\rangle ^{2}}\,.
\end{align}
Actually, only the special case $\xi_{R^{\prime}}^{2}=\frac{N\left(\Delta J_x\right)^{2}}{0\left\langle J_y\right\rangle ^{2}+\left\langle J_z\right\rangle ^{2}}$ is considered in Ref. \cite{DrivenSuperradSpinPrecedent}; but this is not really a loss of generality orientation, as this special orientation is the optimal one for Dicke model driven superradiance. The variance is minimized along $\hat{x}$, as we have seen. When the vectors are chosen accordingly, $\xi_{R^{\prime}}^2$ reduces to $\xi_{R}^2$, for which it is known that ${\xi^2_S} \leq \xi_{R}^2$; see the note following Eq. (68) in Ref. \cite{NoriSpinSqueezing}.  We now show that this inequality is {\em not saturated} when considering Dicke model driven superradiance. 

From the definition of $\jbar_{x/z/y}$ in Ref. \citep[Eqs. (3,35)]{SpinSqueeze2013} we obtained \eq{eg:expecvalueintro}. Using the conventional definitions 
\begin{align}
J_{x/z/y}\equiv \frac{\sigma_{x/y/z}}{2} \otimes \underbrace{\id ... \id}_{N-1}+\id\otimes \frac{\sigma_{x/y/z}}{2} \otimes \underbrace{\id ... \id}_{N-2}\cdots
\end{align}
one obtains that for symmetric states that
\begin{align}
\Braket{J_{x/z/y}}=\frac{N^2\Braket{\sigma_{x/y/z}\;\rho_N^{(1)}}^2}{4}\quad\text{and}\quad \frac{N+N(N-1)\Braket{\sigma_{x/y/z}\otimes\sigma_{x/y/z}\;\rho_N^{(2)}}^2}{4}
\end{align}
which recalling $\left(\Delta J_x\right)^{2}= \Braket{J_x^2}-\Braket{J_x}^2$ leads us to 
\begin{align}
\xi_{R^{\prime}{_N}}^{2}=\frac{1-N\Braket{\sigma_x\;\rho_N^{(1)}}+\left(N-1\right)\Braket{\sigma_x \otimes \sigma_x \;\rho_N^{(2)}}}{\Braket{\sigma_y\;\rho_N^{(1)}}^2+\Braket{\sigma_z\;\rho_N^{(1)}}^2}
\end{align}
which even if we substitute $\Braket{\sigma_{x}\;\rho_N^{(1)}}\to 0$ per \eq{eq:nosigmax} we still obtain a parameter $\xi_{R^{\prime}{_N}}^{2}$ which is unequivocally inequivalent from the alternative parameter in \eq{eq:assumedmin} which defines
\begin{align}\label{eq:ourchisummary}
\xi_{S{_N}}^{2}=1+2\left(N-1\right)\Braket{\sigma_x \otimes \sigma_x \;\rho_N^{(2)}}\,,
\end{align}
see \citep[Table 1]{NoriSpinSqueezing}. Since generally $|\Braket{\sigma_y}^2+\Braket{\sigma_z}^2|<1\,$, we have that $\xi_{R^{\prime}{_N}}^{2}>\xi_{{S}{_N}}^{2}$, and thus we find instances when $\xi_{{S}{_N}}^{2}<1$ but $\xi_{R^{\prime}{_N}}^{2}>1$, indicating entanglement detecting by our criterion by not by $\xi_{R^{\prime}}^{2}>1\,$.

We would like to point out that for states per \eq{eq:rhodef}, our spin squeezing parameter is effectively equivalent to that of \citep[Eq. (7)]{OptimalSpinSqueezingParamater}, listed in Refs. \citep[Eqs. (83-85)]{NoriSpinSqueezing} with subscript $E$ to indicate its optimality at detecting entanglement. This third parameterization also appears earlier in Refs. \citep[Eq. (7c)]{PhysRevLett.99.250405} and \citep[Eq. (7c)]{TothSpinSqueezing}. It is defined as
\begin{align}\label{eq:tothversion}
\xi_{E}^{2}=\frac{(N-1)\left(\Delta J_x\right)^{2}}{
\Braket{J_y^{2}}+\Braket{J_z^{2}}-\frac{N}{2}}
\end{align}
which for symmetric states reduces to
\begin{align}
\xi_{E{_N}}^{2}=\frac{1-N\Braket{\sigma_x\;\rho_N^{(1)}}+\left(N-1\right)\Braket{\sigma_x \otimes \sigma_x \;\rho_N^{(2)}}}{\Braket{\sigma_y \otimes \sigma_y\;\rho_N^{(2)}}+\Braket{\sigma_z \otimes \sigma_z \;\rho_N^{(2)}}}
\end{align}
where we have used $\Braket{\sigma_{x/y/z}\;\rho_N^{(1)}}$ per \eq{eq:nosigmax}. We can simplify further, however, by noting that 
\begin{align}
\Braket{\left(\sigma_x \otimes \sigma_x+\sigma_y \otimes \sigma_y+\sigma_z \otimes \sigma_z\right) \;\rho_N^{(2)}}=\nmap{\Braket{\left(\sigma_x \otimes \sigma_x+\sigma_y \otimes \sigma_y+\sigma_z \otimes \sigma_z\right)\;\rho_2^{\vphantom{(2)}}}}=\nmapalt{\big(1\big)}=1
\end{align}
where we used the trace condition given in \eq{eq:tracecondition}. Additionally substituting $\Braket{\sigma_{x}\;\rho_N^{(1)}}\to 0$ per \eq{eq:nosigmax} we identify
\begin{align}
\xi_{E{_N}}^{2}=\frac{1+\left(N-1\right)\Braket{\sigma_x \otimes \sigma_x \;\rho_N^{(2)}}}{1-\Braket{\sigma_x \otimes \sigma_x \;\rho_N^{(2)}}}
\end{align}
which {\em is} effectively equivalent to the criterion of \eq{eq:ourchisummary} in two important senses:
\begin{enumerate}
\item $\xi_{E{_N}}^{2}\to 1\,$ precisely whenever $\xi_{S{_N}}^{2} \to 1\,$, ie. whenever $\Braket{\sigma_x \otimes \sigma_x \;\rho_N^{(2)}} \to 0\,$. Thus they are equivalent criteria with respect to separability.
\item Both $\xi_{E{_N}}^{2}$ and $\xi_{S{_N}}^{2}$ are pure monotonically-increasing functions of $\Braket{\sigma_x \otimes \sigma_x \;\rho_N^{(2)}}$. Note that this monotonicity only hold in the eigenspectrum of $\Braket{\sigma_x \otimes \sigma_x \;\rho_N^{(2)}}$, that is in the range $-1\leq \Braket{\sigma_x \otimes \sigma_x \;\rho_N^{(2)}}\leq 1\,$, but this comprises all physical observables. This common monotonicity implies that both $\xi_{E{_N}}^{2}$ and $\xi_{S{_N}}^{2}$ are {\em minimized} by the same value of $\Braket{\sigma_x \otimes \sigma_x}$. 
\end{enumerate}
%Note that perfect entanglement in which $\xi_{N}^{2}\to 0$, corresponding to $ \Braket{\sigma_x \otimes \sigma_x \;\rho_N^{(2)}} \to -(n-1)^{-1}$, cannot be achieved in the steady state of driven Dicke Model superradiance; see \fig{fig:figXvN}.

The {\em nature} of the entanglement we observe is also rather curious, in that there appears to be an absence of W-type \cite{PhysRevA.62.062314,PhysRevLett.87.040401,PhysRevA.67.012108,PhysRevA.74.062310,PhysRevA.81.052315} entanglement, which is the form of entanglement possessed by the Dicke basis states. This type of entanglement is recognized as maximal by the geometric measure of entanglement \cite{PhysRevA.82.012327,PhysRevA.80.052315,PhysRevA.82.032301}, although not by Concurrence or Negativity \cite{PhysRevA.70.022322,PhysRevLett.95.260502,PhysRevA.82.012327}. Indeed, the spin squeezing parameter $\xi_{R^{\prime}}^{2}$ of \eq{eq:xiconventional} fails to detect pure Dicke states as entangled at all \cite{PhysRev.93.99,TothDickeArXiv}. Nevertheless, the more sensitive parameter $\xi_{E}^{2}$ of \eq{eq:tothversion} {\em can} detect the entanglement of Dicke states when the variance is taken to be along the $\hat{z}$ direction \citep[Sec. III.C]{OptimalSpinSqueezingParamater}.

When we implement the coordinate-independent basis of  $\xi_{E}^{2}$ \eq{eq:tothversion} \citep[Sec. VI.B)]{NoriSpinSqueezing,OptimalSpinSqueezingParamater}, however, we find that $\xi_{E}^{2}$ is optimally minimized in our steady state system by setting its direction to $\hat{x}$, entirely orthogonal to $\hat{z}$! In this orientation, even $\xi_{E}^{2}$ is blind to the entanglement of the Dicke states. Indeed, that $\xi^2_{_N}\geq 1$ (and therefore  $\xi^2_{E}\geq 1$) for any pure Dicke state is readily apparent in \eq{eq:altchi}: Pick any particular choice for $m^*$ and define 
\begin{align}
\operatorname{X(j)}\limits_{m_a}^{m_b}=\begin{cases}{\binom{2j}{j+m^*}}^{-1} & \text{if }\;m_a=m_b=m^* \\ 0 & \text{else} \end{cases}\,,
%\nicefrac{\bdelta{m_a=m_b=m}}{\binom{2j}{j+m}}\,
\end{align} which is as large as it can be per \eq{eq:tracecondition}. Finally, note that the sum terms in \eq{eq:altchi} are therefore zero for $s>0$ and positive for $s=0$. 
%footnote{\twocolumngrid For an entangled pure Dicke state $\xi^2_{_N}\geq 1$ even though Negativity is nonzero; this does not violate corollary \noeq{eq:negrelation}.} 

Our observation of spin squeezing in the driven superradiating model is therefore all the more interesting; the observed entanglement is entirely distinct from the entanglement of the basis states. 

\section{Tabulation of steady state spin squeezing for various N}
\vspace*{-4ex}
\begin{table}[ht]
\caption{\label{tab:smallspin} $\lim\limits_{t \to \infty}\xi^2_{_N}$. We tabulate the spin-squeezing parameter $\xi^2$ for a steady-state driven-superradiant system for various particle number $N$. $\xi^2_{_N}$ is a continuous function of the driving-frequency to relaxation-frequency ratio, $\Omega=\omega/\Gamma$. Note that \fig{fig:figOvN} in the main text includes plotted curves for $\xi^2_{_N}$ vs $N$, for $N=2,4,8$ analytical expressions for which appear in this table.}
\begin{ruledtabular}
\begin{tabular}{ccl}
 2 && \(1-\Omega ^2\frac{ 2-\Omega ^2}{3 \Omega ^4+4 \Omega ^2+4}\) \\
 3 && \(1-\Omega ^2\frac{ -4 \Omega ^4+10 \Omega ^2+12}{6 \Omega ^6+15 \Omega ^4+36 \Omega ^2+54}\) \\
 4 && \(1-\Omega ^2\frac{ -5 \Omega ^6+15 \Omega ^4+42 \Omega ^2+72}{5 \Omega ^8+20 \Omega ^6+84 \Omega ^4+288 \Omega ^2+576}\) \\
 5 && \(1-\Omega ^2\frac{ -40 \Omega ^8+140 \Omega ^6+672 \Omega ^4+2592 \Omega ^2+5760}{30 \Omega ^{10}+175 \Omega ^8+1120 \Omega ^6+6480 \Omega ^4+28800 \Omega ^2+72000}\) \\
 6 && \(1-\Omega ^2\frac{ -35 \Omega ^{10}+140 \Omega ^8+1008 \Omega ^6+6480 \Omega ^4+31680 \Omega ^2+86400}{21 \Omega ^{12}+168 \Omega ^{10}+1512 \Omega ^8+12960 \Omega ^6+95040 \Omega ^4+518400 \Omega ^2+1555200}\) \\
 7 && \(1-\Omega ^2\frac{ -28 \Omega ^{12}+126 \Omega ^{10}+1260 \Omega ^8+11880 \Omega ^6+95040 \Omega ^4+561600 \Omega ^2+1814400}{14 \Omega ^{14}+147 \Omega ^{12}+1764 \Omega ^{10}+20790 \Omega ^8+221760 \Omega ^6+1965600 \Omega ^4+12700800 \Omega ^2+44452800}\) \\
 8 && \(1-\Omega ^2\frac{ -7 \Omega ^{14}+35 \Omega ^{12}+462 \Omega ^{10}+5940 \Omega ^8+68640 \Omega ^6+655200 \Omega ^4+4536000 \Omega ^2+16934400}{3 \Omega ^{16}+40 \Omega ^{14}+616 \Omega ^{12}+9504 \Omega ^{10}+137280 \Omega ^8+1747200 \Omega ^6+18144000 \Omega ^4+135475200 \Omega ^2+541900800}\) \\
 9 && \(1-\Omega ^2\frac{ -80 \Omega ^{16}+440 \Omega ^{14}+7392 \Omega ^{12}+123552 \Omega ^{10}+1921920 \Omega ^8+26208000 \Omega ^6+290304000 \Omega ^4+2303078400 \Omega ^2+9754214400}{30 \Omega ^{18}+495 \Omega ^{16}+9504 \Omega ^{14}+185328 \Omega ^{12}+3459456 \Omega ^{10}+58968000 \Omega ^8+870912000 \Omega ^6+10363852800 \Omega ^4+87787929600 \Omega ^2+395045683200}\) \\

\end{tabular}
\end{ruledtabular}
\end{table}
\vspace*{-1ex}
\putbib[NoURL]
\end{bibunit}

\end{document}